\documentclass[journal,onecolumn,draftcls]{IEEEtran}
\usepackage{url}
\usepackage{ifpdf}
\ifpdf
\usepackage[pdftex]{graphicx}
\else
\usepackage[dvips]{graphicx}
\fi

\DeclareGraphicsExtensions{.eps}
\usepackage{epstopdf}
\usepackage{tikz}

\usepackage{caption}
\usepackage[caption=false]{subfig}
\usepackage{epsfig}
\usepackage{amsmath}
\usepackage{url}
\usepackage{color}
\usepackage{algorithm}
\usepackage{algpseudocode}
\usepackage{mathcomp}
\usepackage{mathtools}
\usepackage{amssymb}
\usepackage{amsmath}
\usepackage{amsthm}
\usepackage{multicol}
\usepackage{multirow}

\newtheorem*{rep@theorem}{\rep@title}
\newcommand{\newreptheorem}[2]{%
	\newenvironment{rep#1}[1]{%
		\def\rep@title{#2 \ref{##1}}%
		\begin{rep@theorem}}%
		{\end{rep@theorem}}}
\makeatother
\newreptheorem{theorem}{Theorem}

\newtheorem{lemma}{Lemma}

\usepackage{mathrsfs}
\usepackage{graphicx}
\usepackage{hhline}
\usepackage{bigstrut}
\usepackage{cases}
\usepackage{verbatim}
\usepackage{bm}
\usepackage[numbers,sort&compress]{natbib}

\begin{document}
\title{Energy Storage Arbitrage in Real-Time Markets via Reinforcement Learning}

\author{
\IEEEauthorblockN{Hao~Wang,~Baosen~Zhang}\\
\IEEEauthorblockA{Department of Electrical Engineering, 
University of Washington,
Seattle, WA 98195\\
Email: \{hwang16,zhangbao\}@uw.edu}
}


%Telephone: (800) 555--1212, Fax: (888) 555--1212}


\maketitle

\begin{abstract}
In this paper, we derive a temporal arbitrage policy for storage via reinforcement learning. Real-time price arbitrage is an important source of revenue for storage units, but designing good strategies have proven to be difficult because of the highly uncertain nature of the prices. Instead of current model predictive or dynamic programming approaches, we use reinforcement learning to design an optimal arbitrage policy. This policy is learned through repeated charge and discharge actions performed by the storage unit through updating a value matrix. We design a reward function that does not only reflect the instant profit of charge/discharge decisions but also incorporate the history information. Simulation results demonstrate that our designed reward function leads to significant performance improvement compared with existing algorithms.

\end{abstract}

\IEEEpeerreviewmaketitle

\section{Introduction}\label{intro}
%% !TEX root=main.tex

Energy storage can provide various services (e.g., load shifting, energy management, frequency regulation, and grid stabilization) \cite{eyer2010energy} to the power grid and its economic viability is receiving increasing attention. One of the most discussed revenue sources for energy storage is \emph{real-time temporal arbitrage}~(i.e., charging at low prices and discharging at higher prices), where storage units take advantage of the price spreads in real-time electricity market prices to make profits over time~\cite{sioshansi2009estimating}. This application has received significant attention from the research community, especially since the growing penetration of intermittent renewable generations are resulting in more volatile real-time electricity market prices~\cite{woo2011impact}. However, even with this increase in price spread, it remains nontrivial to design arbitrage policies that make significant (or even positive) profit~\cite{byrne2012estimating}. The difficulties come from the fact that future prices are unknown, difficult to forecast and may even be non-stationary~\cite{KimEtAl2011,Borenstein2005}.
% difficult and uncertain prices the future prices are unknown, difficult to forecast and potentially it is challenging to determine when to design and how to operate energy storage in real-time markets is not well-understood.
In this paper, we aim to develop an arbitrage policy for energy storage in a data-driven framework by using reinforcement learning~\cite{SuttonEtAl2011}.

For example, arbitrage using energy storage has been studied in~\cite{walawalkar2007economics,sioshansi2009estimating,bradbury2014economic,mcconnell2015estimating,zafirakis2016value}~(and see the references within). The authors in \cite{walawalkar2007economics} studied using sodium-sulfur batteries and flywheels for arbitrage in NYISO found the batteries can be potentially profitable using data from 2001 to 2004. The authors in \cite{sioshansi2009estimating} analyzed a generic storage system in the PJM real-time market and discovered that the arbitrage value was nearly doubled from 2002 to 2007 due to higher price variations. The authors in \cite{bradbury2014economic} formulated a linear optimization problem to compare the arbitrage profits of 14 energy storage technologies in several major U.S. real-time electric markets. Similar studies have also been carried out in different markets, e.g., Australian national electricity market \cite{mcconnell2015estimating} and European electricity markets \cite{zafirakis2016value}. Crucially, all of these studies assumed perfect knowledge of electricity prices and therefore cannot be implemented as real-time arbitrage strategies.

Some recent works~\cite{abdulla2016optimal,krishnamurthy2017energy,jiang2015optimal,qin2016online} have started to explicitly take the electricity price uncertainty into account when designing arbitrage strategies.  The authors in \cite{abdulla2016optimal} proposed a stochastic dynamic program to optimally operate an energy storage system using available forecast. The authors in \cite{krishnamurthy2017energy} formulated a stochastic optimization problem for a storage owner to maximize the arbitrage profit under uncertainty of market prices. Both studies need to forecast electricity prices and their performances heavily rely on the quality of the forecast. However, real-time market prices are highly stochastic and notoriously difficult to forecast well~\cite{Weron2014}. To overcome the reliance on price predictions, the authors in \cite{jiang2015optimal} employed approximate dynamic programming to derive biding strategy for energy storage in NYISO real-time market without requiring prior knowledge of the price distribution. However, this strategy is often highly computationally expensive. The authors in \cite{qin2016online} proposed an online modified greedy algorithm for arbitrage which is computationally straightforward to implement and does not require the full knowledge of price distributions. But it needs to estimate the bounds of prices and assume that storages are ``big enough'', which is not always true in practice.

The aforementioned challenges motivate us to develop an easily implementable arbitrage policy using \emph{reinforcement learning (RL)}. This policy is both price-distribution-free and outperforms existing ones. Without explicitly assuming a distribution, our policy is able to operate under constantly changing prices that may be non-stationary. Over time, by repeatedly performing charge and discharge actions under different real-time prices, the proposed RL-based policy learns the best strategy that maximizes the cumulative reward. The key technical challenge turns out to be the design of a good reward function that will guide the storage to make the correct decisions. Specifically, we make the following two contributions in this paper:
\begin{enumerate}
	\item We formulate the energy storage operation as a Markov decision process (MDP) and derive a Q-learning policy~\cite{watkins1992q} to optimally control the charge/discharge of the energy storage for temporal arbitrage in the real-time market.
	\item We design a reward function that does not only reflect the instant profit of charge/discharge decisions but also incorporate historical information. Simulation results demonstrate that the designed reward function leads to significant performance improvements compared to the natural instant reward function. In addition, using real historical data, we show the proposed algorithm also leads to much higher profits than existing algorithms.
\end{enumerate}
 % In this paper, we formulate the energy arbitrage problem as a Markov decision process (MDP) and design an arbitrage algorithm via reinforcement learning. Different from the thresholding strategy used in \cite{qin2016online}, our algorithm does not force any specific strategies but learns the best strategy to decide when to charge and discharge to maximize the profit.

%\updated{Some of the key findings are summarized as follows. }

The remainder of the paper is ordered as follows. In Section~\ref{model}, we present the optimization problem for energy storage arbitrage. In Section~\ref{solution}, we provide a reinforcement learning approach to obtain the arbitrage policy. Numerical simulations using real data are discussed in Section~\ref{simulation}. Section~\ref{conclusion} concludes this paper.

\section{Arbitrage Model and Optimization Problem}\label{model}
%% !TEX root=main.tex

We consider an energy storage (e.g., a battery) operating in a real-time electricity market over a finite operational horizon $\mathcal{T} = \{1,...,T\}$. The objective of the energy storage is to maximize its arbitrage profit by charging at low prices and discharging when prices are high. We assume the energy storage is a price taker, and its operation will not affect the market prices. We denote $d_t$ as the discharged power from the storage at time $t$ and $c_t$ as the charged power into the storage at time $t$. Let the real-time prices be denoted as $p_{t}$. We formulate the Arbitrage Maximization Problem (AMP) as follows:
\begin{alignat}{2}
\max \quad
& \sum_{t=1}^{T} p_{t} \left( \eta_d d_t - \frac{1}{\eta_c} c_t \right) &{} \tag{AMP} \label{eq-obj} \\
\mbox{subject to} \quad
& E_{t} = E_{t-1} +  c_t  -  d_t, ~ \forall t \in \mathcal{T} \label{eq-con1}\\
& E_{\min} \leq E_{t} \leq E_{\max}, ~ \forall t \in \mathcal{T} \label{eq-con2}\\
& 0 \leq c_t \leq C_{\max},  ~ t \in \mathcal{T} \label{eq-con3}\\
& 0 \leq d_t \leq D_{\max},  ~ t \in \mathcal{T} \label{eq-con4}\\
\mbox{variables:} \quad
& c_t,~d_t,~\forall t \in \mathcal{T}, \nonumber
\end{alignat}
where $\eta_c \in (0,1)$ and $\eta_d \in (0,1)$ denote the charge/discharge efficiencies.
% At time $t$, the charge power from the grid is $\frac{1}{\eta_c} c_t$ and the discharge power into the grid is $\eta_d d_t$.
 The constraint in \eqref{eq-con1} specifies the dynamics of energy level $E_{t}$ over time, \eqref{eq-con2} constraints the amount of energy in the storage to be between  $E_{\min}$ and $E_{\max}$, \eqref{eq-con3} and \eqref{eq-con4} bounds the maximum charge and discharge rates (denoted by $C_{\max}$ and $D_{\max}$, respectively) of the storage.

The optimization problem in AMP is a linear program, and we characterize its optimal solution in the next lemma.
\begin{lemma}\label{optsol}
	The optimal charge and discharge profiles $\{c_t^{\star},~d_t^{\star},~\forall t \in \mathcal{T}\}$ satisfy
		\begin{enumerate}
		\item At least one of $c_t^{\star}$ or $ d_t^{\star}$ is $0$ at any time $t$;
		\item $c_t^{\star} = \{ 0, \min \{C_{\max}, E_{\max}-E_{t-1}\} \}$,\\ $d_t^{\star} = \{ 0, \min \{ D_{\max}, E_{t-1}-E_{\min}\} \}$.
		\end{enumerate}
\end{lemma}
Lemma \ref{optsol} states that the energy storage will not charge and discharge at the same time. Also, the optimal charge and discharge power will hit the boundary per the operational constraints \eqref{eq-con1}-\eqref{eq-con4}. Specifically, when the storage decides to charge, it will charge either at the maximum charge rate $C_{\max}$ or reaching the maximum energy level $E_{\max}$. Similarly, the discharge power will be either the maximum discharge rate $D_{\max}$ or the amount to reach the minimum energy level $E_{\min}$. This binary charging/discharging structure will be important when we design the reinforcement learning algorithm in the next section.

If the future prices are known, the optimization problem in AMP can be easily solved to provide an offline optimal strategy for the charge/discharge decisions. However, the offline solution is only practical if a good price forecast is available. In reality, future prices are not known in advance and the energy storage needs to make decisions based on only the current and historical data. In other words, the charge/discharge decisions $\{\hat{c}_t,~\hat{d}_t\}$ are functions of price information up to the current time slot $t$, denoted by $\{ p_1, ..., p_t \}$:
\begin{equation}
 \{\hat{c}_t,~\hat{d}_t\} = \pi (p_1, ..., p_t), \label{policy}
\end{equation}
where $\pi(\cdot)$ is the arbitrage policy for maximizing the profit. Therefore, AMP is a constrained sequential decision problem and can be solved by dynamic programming~\cite{howard1960dynamic}. But the potentially high dimensionality the state space makes dynamic programming computationally expensive, and potentially unsuitable for applications like real-time price arbitrage. Moreover, price forecast in real-time markets is extremely challenging, as the mismatch between power supply and demand can be attributed to many different causes.
% , e.g., uncertainty in renewable dispatch, load prediction errors, and congestions. To overcome the aforementioned shortcomings, in Section \ref{solution}, we will formulate AMP as an MDP and derive an effective arbitrage policy via reinforcement learning.}

\section{Reinforcement Learning Algorithm}\label{solution}
%% !TEX root=main.tex
To solve the \emph{online version} of AMP, we use reinforcement learning (RL).
Reinforcement learning is a general framework to solve problems in which~\cite{SuttonEtAl2011}: (i) actions taken depend on the system states; (ii) a cumulative reward is optimized; (iii) only the current state and past actions are known; (iv) the system might be non-stationary. The energy storage arbitrage problem has all of the four properties: (i) different electricity prices lead to different actions (e.g., charge/discharge), and the future energy storage level depends on past actions; (ii) the energy storage aims at maximizing the total arbitrage profit; (iii) the energy storage does not have a priori knowledge of the prices, while it knows the past history; (iv) the actual price profiles are non-stationary. In the following, we describe the RL setup for AMP in more detail.

% \updated{As discussed in Section \ref{model}, AMP is a sequential decision problem and we model it as an MDP, which consists of state, action, reward, and transition as follows.}

\subsection{State Space}
We define the state space of the energy arbitrage problem as a finite number of states. To be specific, the system's state can be fully described by the current price $p_t$ and previous energy level $E_{t-1}$. We discretize the price into $M$ intervals, and energy level into $N$ intervals. %where $N = \lceil \frac{ E_{\max}-E_{\min} }{ \min\{C_{\max},D_{\max}\}}\rceil$, such that the state space is given by
\begin{equation*}
\mathcal{S} = \{ 1, ..., M \} \times \{1,..., N\},
\end{equation*}
where $\{ 1, ..., M \}$ represents $M$ even price intervals from the lowest to the highest, and $\{1,..., N\}$ denotes $N$ energy level intervals ranging from $E_{\min}$ to $E_{\max}$.

\subsection{Action Space}
Per Lemma \ref{optsol}, the energy storage will not charge and discharge at the same time. Moreover, the optimal charge and discharge power always reach their maximum allowable rates. We denote the maximum allowable charge/discharge rates as $\tilde{D}_{\max} = \min \{ D_{\max}, E_{t-1}-E_{\min}\}$ and $\tilde{C}_{\max} = \min \{C_{\max}, E_{\max}-E_{t-1}\}$. Therefore, the action space of the energy storage consists of three actions: charge at full rate, hold on, and discharge at full rate:
\begin{equation*}
\mathcal{A} = \{ -\tilde{D}_{\max}, 0, \tilde{C}_{\max} \},
\end{equation*}
where action $a = -\tilde{D}_{\max}$ denotes discharge either at maximum rate $D_{\max}$ or unitl the storage hits the minimum level $E_{\min}$. Action $a = \tilde{C}_{\max}$ denotes charge at maximum rate $C_{\max}$ or until the storage reaches the maximum level $E_{\max}$.

\subsection{Reward}
At time $t$, after taking an action $a \in \mathcal{A}$ at state $s \in \mathcal{S}$, the energy storage will receive a \emph{reward}, such that the energy storage knows how good its action is. According to the objective function of AMP, the energy storage aims to maximize the arbitrage profit by charging at low prices and discharge at high prices. Therefore, we can define the reward as
\begin{alignat}{2}
r_t^{1} =
\begin{cases}
-p_t \tilde{C}_{\max}  & \mbox{if charge} \\
0 & \mbox{if hold on} \\
p_t \tilde{D}_{\max}  & \mbox{if discharge} \tag{Reward 1}
\end{cases}
\end{alignat}
which is the instant reward of charge or discharge. If the energy storage charges at the rate of $\tilde{C}_{\max}$ at time $t$, it will pay at the spot price and reward is negative, i.e., $-p_t \tilde{C}_{\max}$. In contrast, the energy storage discharges at the rate of $\tilde{D}_{\max}$ and will earn a revenue of $p_t \tilde{D}_{\max}$.

Reward 1 is a straightforward and natural design, but is actually not very effective. The reason is that the negative reward for charge makes the energy storage perform conservatively in the learning process and thus the arbitrage opportunity is under explored. This motivates us to develop a more effective reward. To avoid conservative actions, we introduce an average price in the reward. The idea comes from the basic principle of arbitrage: to charge at low prices and discharge at high prices. The average price works as a simple indicator to determine whether the current price is low or high \emph{compared to the historical values}. Specifically, the new reward is defined as
\begin{alignat}{2}
r_t^{2} =
\begin{cases}
(\overline{p}_t - p_t) \tilde{C}_{\max}  & \mbox{if charge} \\
0 & \mbox{if hold on} \\
(p_t - \overline{p}_t) \tilde{D}_{\max}  & \mbox{if discharge} \tag{Reward 2}
\end{cases}
\end{alignat}
where the average price $ \overline{p}_t$ is calculated by
\begin{equation}
 \overline{p}_t = (1-\eta) \overline{p}_{t-1} + \eta p_t, \label{ave}
\end{equation}
in which $\eta$ is the smoothing parameter. Note that $ \overline{p}_t$ is not a simple average that weighs all past prices equally. Instead, we use moving average in \eqref{ave}, such that we not only leverage the past price information but also adapt to the current price change.

We see from Reward 2 that when the energy storage charges at a price lower than the average price (i.e., ${p}_t  < \overline{p}_t$), it will get a positive reward $(\overline{p}_t - p_t) \tilde{C}_{\max} > 0$, otherwise it will receive a loss if the spot price is greater. Similarly, Reward 2 encourages the energy storage to discharge at high price by giving a positive reward, i.e., $(p_t - \overline{p}_t) \tilde{D}_{\max} > 0$. Reward 2 outperforms Reward 1 in exploring more arbitrage opportunities and achieving higher profits. It also mitigates the non-stationarity of prices, since it weights the current price much heavier than prices in the more distant past. We will show the numerical comparisons in Section \ref{simulation}.

\subsection{Q-Learning Algorithm}
With the state, action and reward defined, we obtain the real-time charge and discharge policy using \emph{Q-learning}~(a popular subclass of RL algorithms~\cite{SuttonEtAl2011}). Here the energy storage maintains a state-action value matrix $Q$, where each entry $Q(s,a)$ is defined for each pair of state $s$ and action $a$. When the energy storage takes a charge/discharge action under a spot price, the value matrix is updated as follows:
\begin{equation} \label{qlearn}
Q(s,a)_t = (1-\alpha) Q(s,a)_{t-1} +  \alpha [r_t + \gamma \max_{a'} Q(s',a')],
\end{equation}
where the parameter $\alpha \in (0,1]$ is the learning rate weighting the past value and new reward. $\gamma \in [0,1]$ is the discount rate determining the importance of future rewards. After taking an action $a$, the state transits from $s$ to $s'$, and the energy storage updates the value matrix incorporating the instant reward $r_t$ (e.g., Reward 1 or 2) and the future value $\max_{a'} Q(s',a')$ in state $s'$. Over time, the energy storage can learn the value each action in all states. When $Q(s,a)$ converges to the optimal state-action values, we obtain the optimal arbitrage policy. Specifically, the Q-learning algorithm can derive an arbitrage policy for \eqref{policy} as
\begin{equation} \label{optpolicy}
a^{\star} = \pi(s) = \arg \max_{a} Q(s,a),
\end{equation}
which is the optimal arbitrage policy guranteed for finite MDP \cite{watkins1992q}. For any state $s$, the energy storage always chooses the best action $a^{\star}$ which maximizes the value matrix $Q(s,a)$. 

	\begin{algorithm}[!htbp]
	\caption{Q-learning for energy storage arbitrage}
	\label{alg1}
	\begin{algorithmic}[1]
		\State \textbf{Initialization}: In each time slot $t \in \{1,...,T \}$, set the iteration count $k=1$, $\alpha = 0.5$, $\alpha = 0.9$, and $\epsilon = 0.9$. Initialize the $Q$-matrix, i.e., $Q = 0$.
		\Repeat
		\State \textbf{Step1:} Observe state $s$ based on price and energy level;

		\State \textbf{Step2:} Decide the best action $a$ (using $\epsilon$-greedy method) based on $Q(s,a)$;

		\State \textbf{Step3:} Calculate the reward (using Reward 1 or 2);

		\State \textbf{Step4:} Update $Q(s,a)$ according to \eqref{qlearn} and energy level in \eqref{eq-con1};

		\State $s \gets s'$ and $k \gets k+1$;

		\Until end of operation, i.e., $t=T$.
		\State \textbf{end}
	\end{algorithmic}
\end{algorithm}

The step-by-step Q-learning algorithm for energy arbitrage is presented in Algorithm \ref{alg1}. To avoid the learning algorithm getting stuck at sub-optimal solutions, we employ \emph{$\epsilon$-greedy}~\cite{watkins1992q}. The algorithm not only exploits the best action following \eqref{optpolicy} but also explores other actions, which could be potentially better. Specifically, using $\epsilon$-greedy, the algorithm will randomly choose actions with probability $\epsilon \in [0,1]$, and choose the best action in \eqref{optpolicy} with probability $1-\epsilon$. 

\section{Numerical Results}\label{simulation}
%% !TEX root=main.tex

\begin{figure}[!htb]
	\centering
	\includegraphics[width=0.5\textwidth]{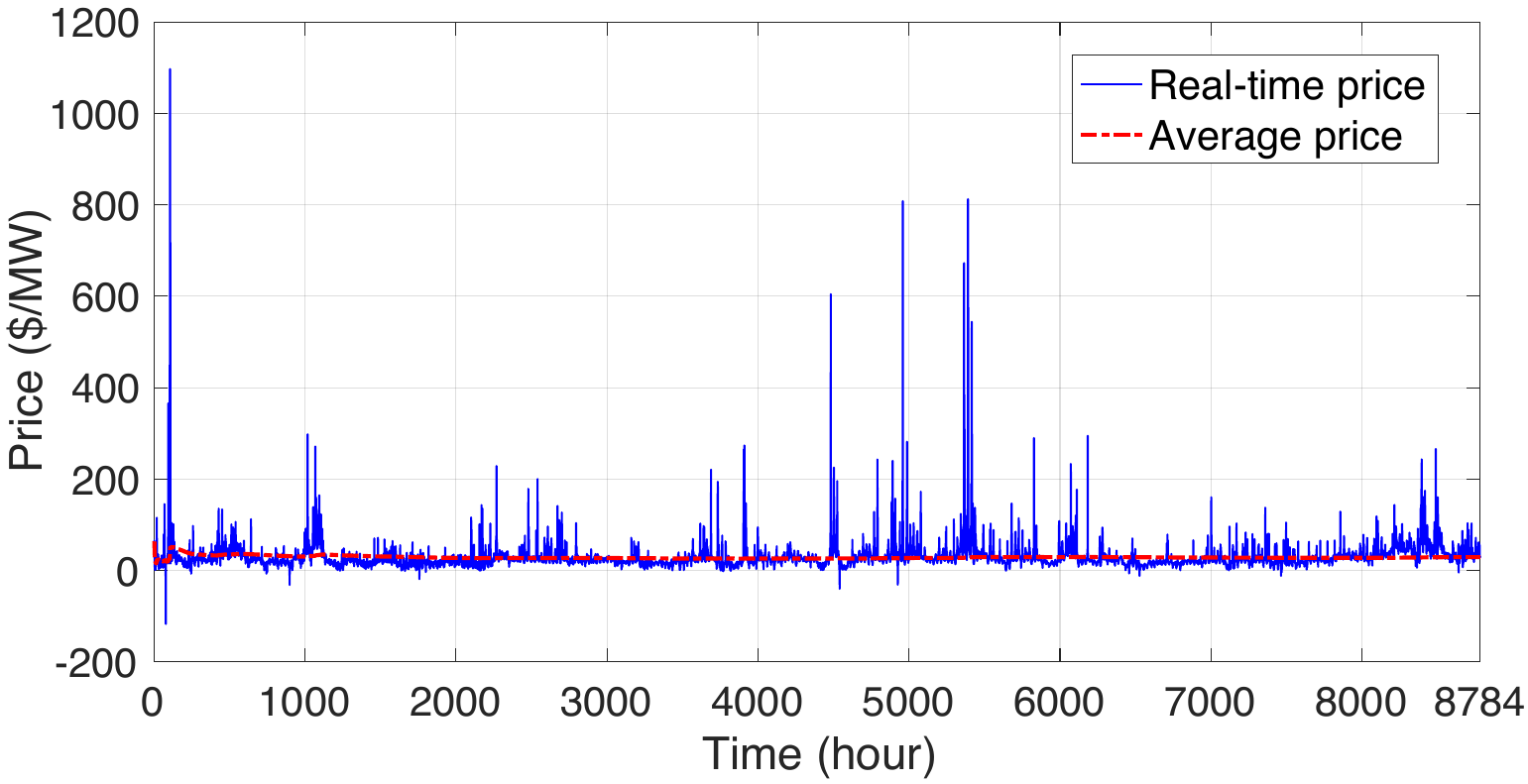}
	\caption{\label{fig_realprice} PJM Real-time price.}
\end{figure}

In this section, we evaluate two reward functions and also compare our algorithm to a baseline in~\cite{qin2016online} under both synthetic prices and realistic prices. For synthetic prices, we generate i.i.d. (independent and identically distributed) prices, and for the realistic price, we use hourly prices from ISO New England real-time market \cite{isonedata} from January 1, 2016 to December 31, 2017.
The realistic price is depicted in Figure \ref{fig_realprice}. We see that the averaged price is flat but the instantaneous prices fluctuate significantly with periodic spikes.

\newpage
\subsection{Synthetic Price}
\begin{figure}[b]%!htb
	\centering
		\includegraphics[width=0.5\textwidth]{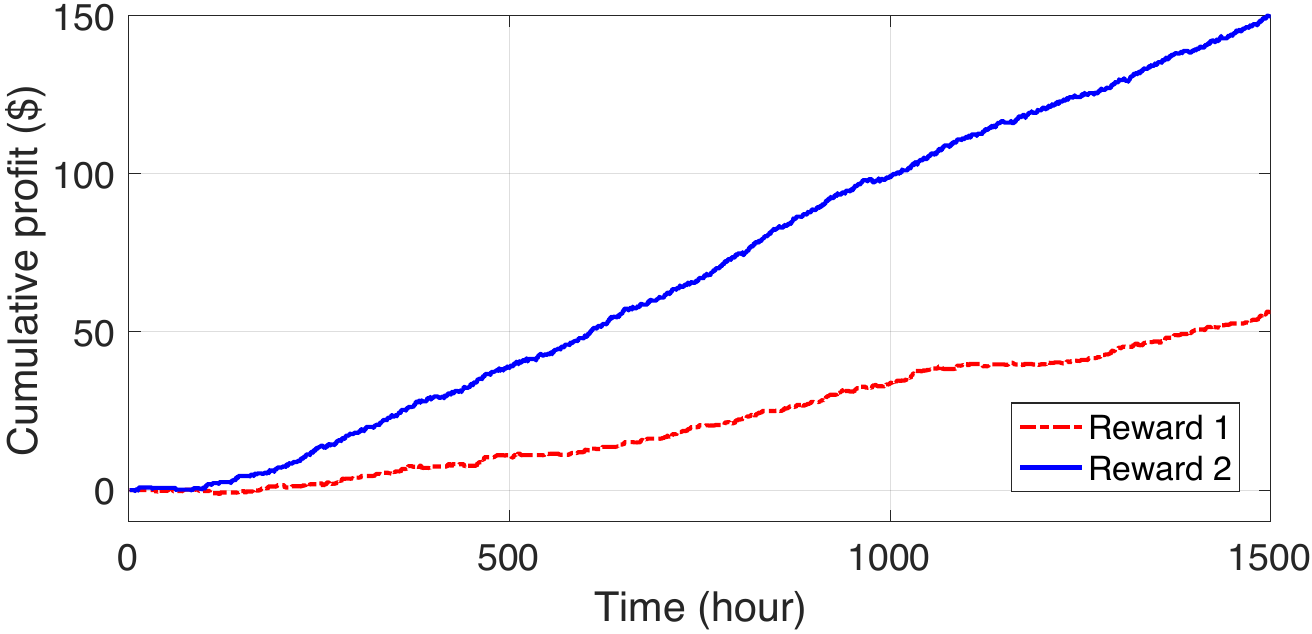}
		\caption{\label{fig_profit_sp} Cumulative profits under synthetic prices.}
\end{figure}

We first evaluate the two reward functions under synthetic price, which is uniformly distributed in $[0,1]$ over $1500$ hours. We set $C_{\max} = D_{\max} =1$, $E_{\min} =0$  and $E_{\max} =1$. The cumulative profits for both rewards are depicted in Figure \ref{fig_profit_sp}. Both profits stay flat over the first 300 hours, as the algorithm is exploring the environment with different prices. Afterwards, the algorithm using Reward 2 starts to make profit and achieves $166\%$ more than Reward 1 in the end.

\begin{figure}
	\begin{minipage}[!htbp]{\linewidth}
		\centering
		\includegraphics[width=0.5\linewidth]{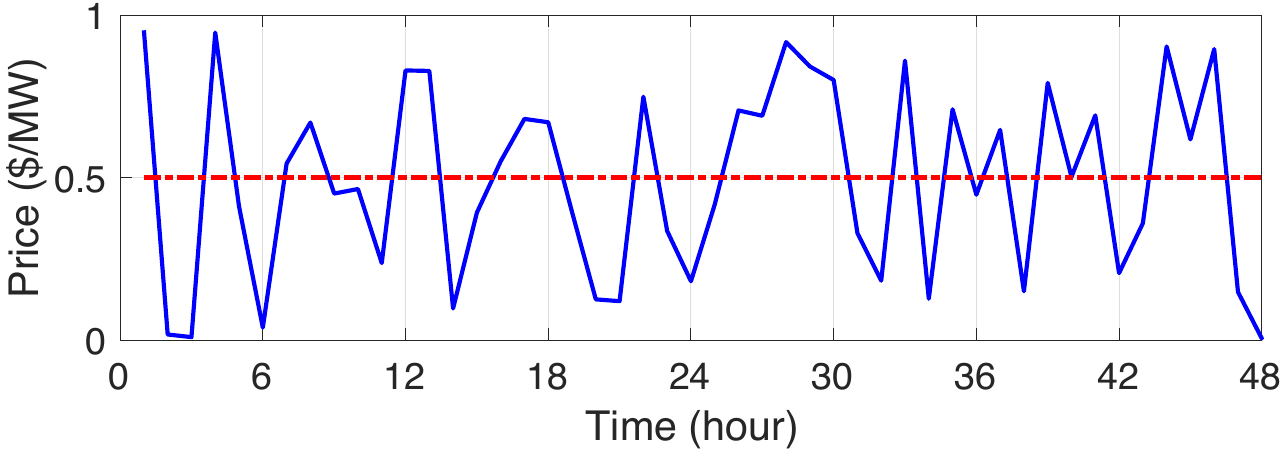}
		\caption*{(a) Synthetic price.}
	\end{minipage}
	\begin{minipage}[!htbp]{\linewidth}
		\centering
		\includegraphics[width=0.5\linewidth]{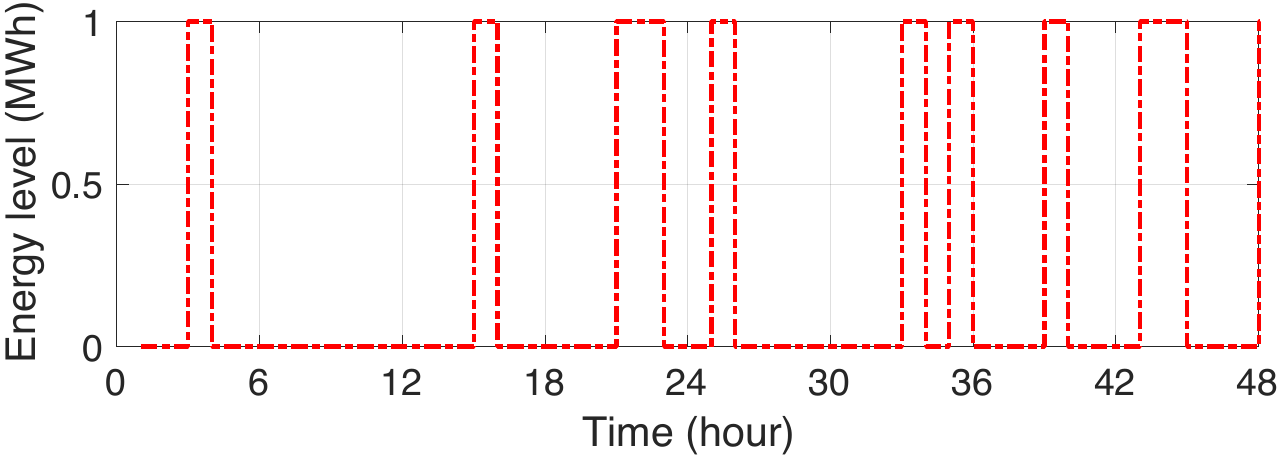}
		\caption*{(b) Energy level of algorithm using Reward 1.}
	\end{minipage}
\begin{minipage}[!htbp]{\linewidth}
	\centering
	\includegraphics[width=0.5\linewidth]{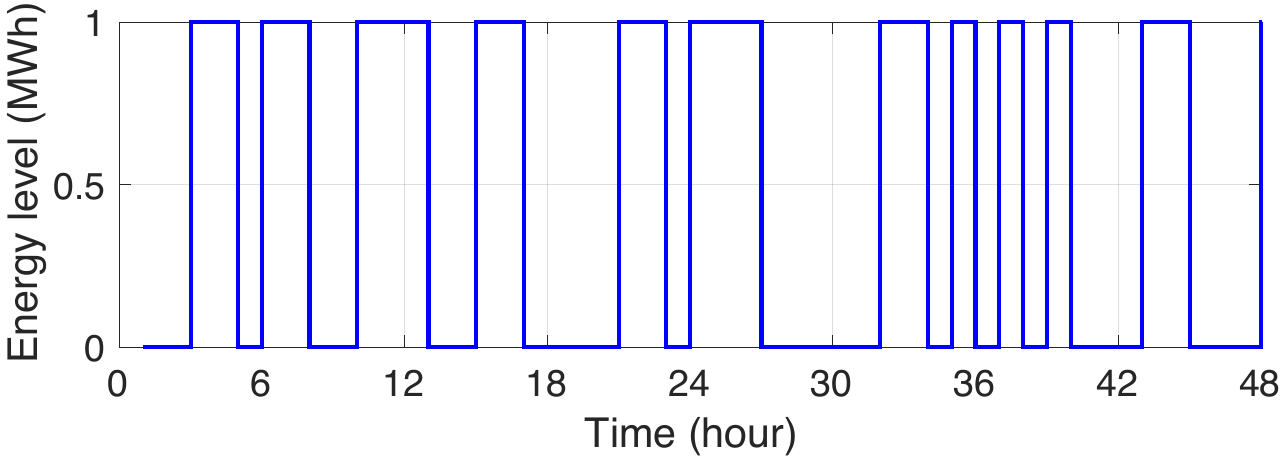}
	\caption*{(c) Energy level of algorithm using Reward 2.}
\end{minipage}
\caption{Price and energy levels over a 48 hour period using reward 1 and reward 2 under synthetic prices.}
\label{fig_profile_sp}
\end{figure}

To further understand the how Reward 1 and Reward 2 affect the storage operation, we plot the evolution of energy level over a 48 hour horizon in Fig. \ref{fig_profile_sp}. We see that algorithm using Reward 1 performs conservatively while Reward 2 makes the algorithm actively charge and discharge to take advantage of price spread. Therefore, Reward 2 leads to a more profitable arbitrage strategy.

\subsection{Real Historical Price}
\begin{figure}[b]%!htb
	\centering
	\includegraphics[width=0.5\textwidth]{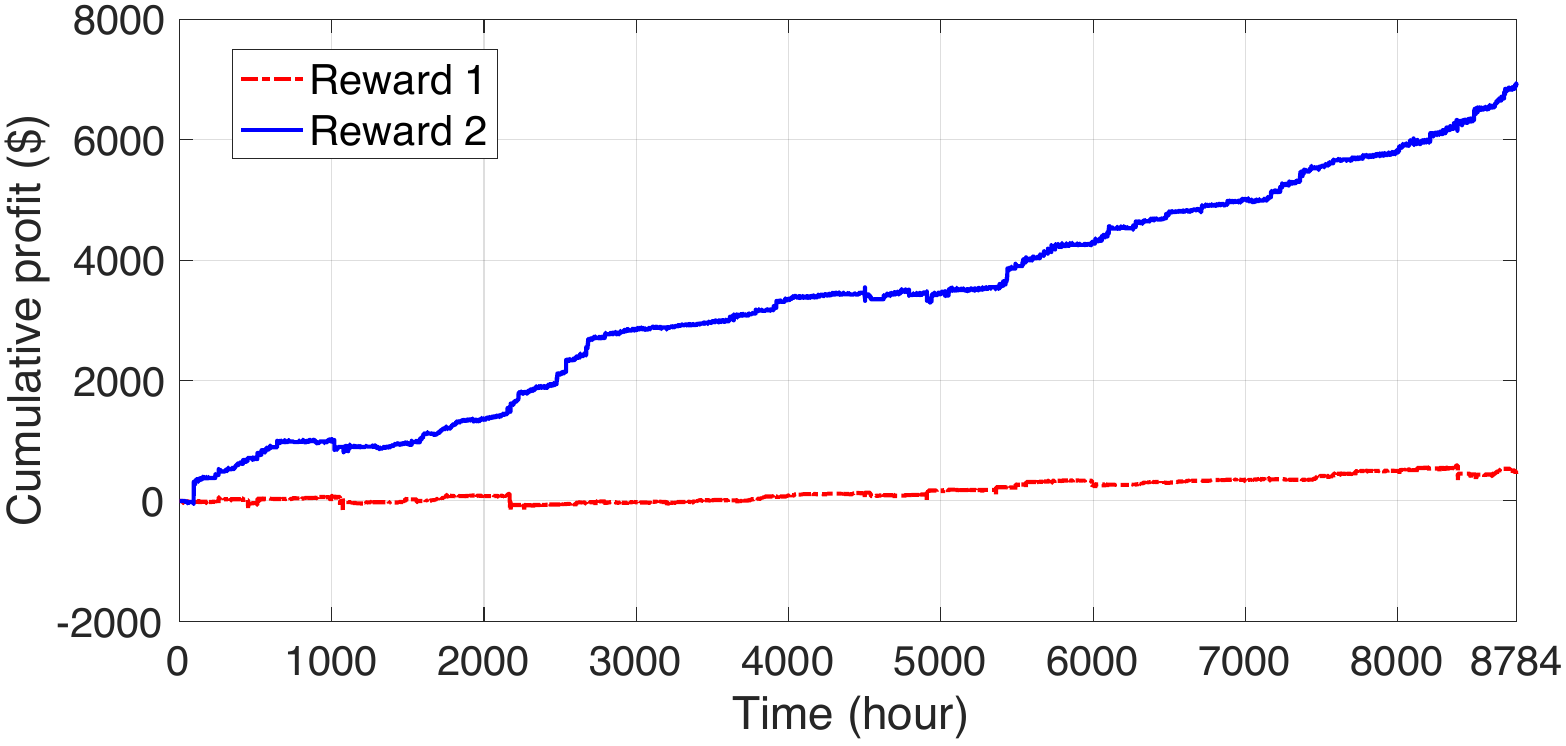}
	\caption{\label{fig_profit_rp} Cumulative profits under real-time prices.}
\end{figure}

We evaluate the two reward functions using realistic prices from ISO New England real-time market in 2016. We plot the cummulative profits of two rewards during training in Figure \ref{fig_profit_rp}. We see that Reward 1 fails to make profit while using Reward 2 produces a high profit. This demonstrates the effectiveness of our designed reward: it is able to adapt to price changes and makes profit continuously.

\begin{figure}
	\begin{minipage}[!htbp]{\linewidth}
		\centering
		\includegraphics[width=0.5\linewidth]{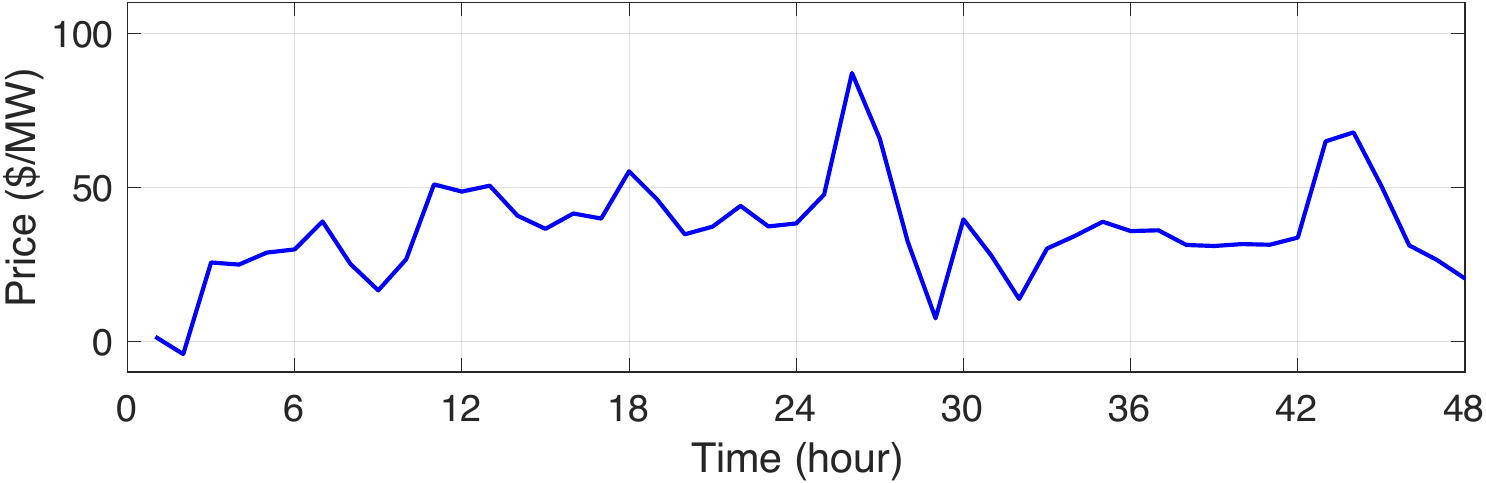}
		\caption*{(a) Real-time price.}
	\end{minipage} \\
	\begin{minipage}[!htbp]{\linewidth}
		\centering
		\includegraphics[width=0.5\linewidth]{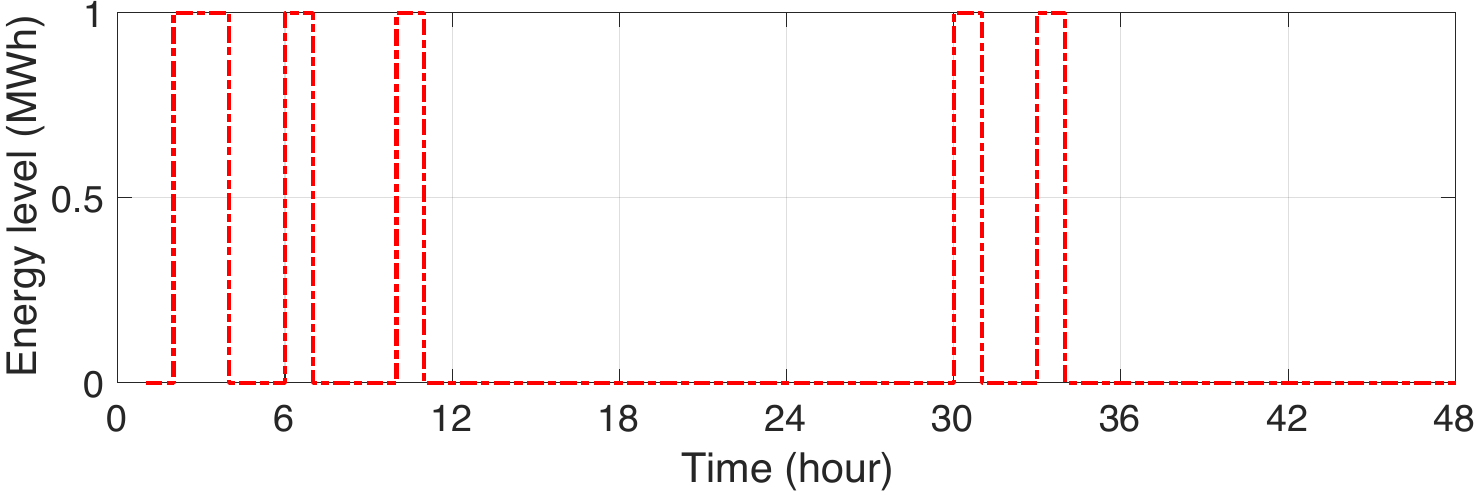}
		\caption*{(b) Energy level of algorithm using Reward 1.}
	\end{minipage} \\
	\begin{minipage}[!htbp]{\linewidth}
		\centering
		\includegraphics[width=0.5\linewidth]{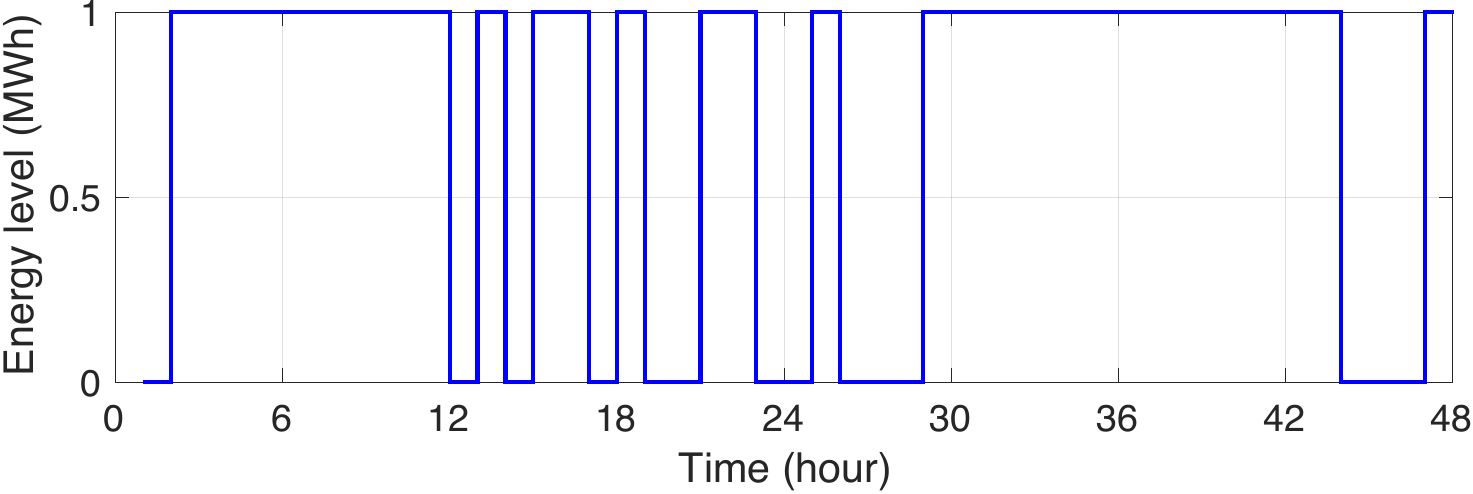}
		\caption*{(c) Energy level of algorithm using Reward 2.}
	\end{minipage}
	\caption{Price and energy levels over a 48 hour horizon for reward 1 and reward 2 under historical data.}
	\label{fig_profile_rp}
\end{figure}

We also plot the evolution of energy levels over a 48-hours operational horizon in Figure \ref{fig_profile_rp}. We see that algorithm using Reward 1 cannot capture the price differences but makes charge/discharge when the real-time price is flat. In contrast, our algorithm using Reward 2 is able to charge at low prices at hours 2 and 29, hold the energy when prices are low, and discharge at hours 12 and 44, respectively, when the price reaches a relatively high point.

\subsection{Comparison with baseline algorithm}
\begin{figure}[!htb]
		\centering
	\begin{minipage}[!htbp]{0.46\linewidth}
		\centering
		\includegraphics[width=\linewidth]{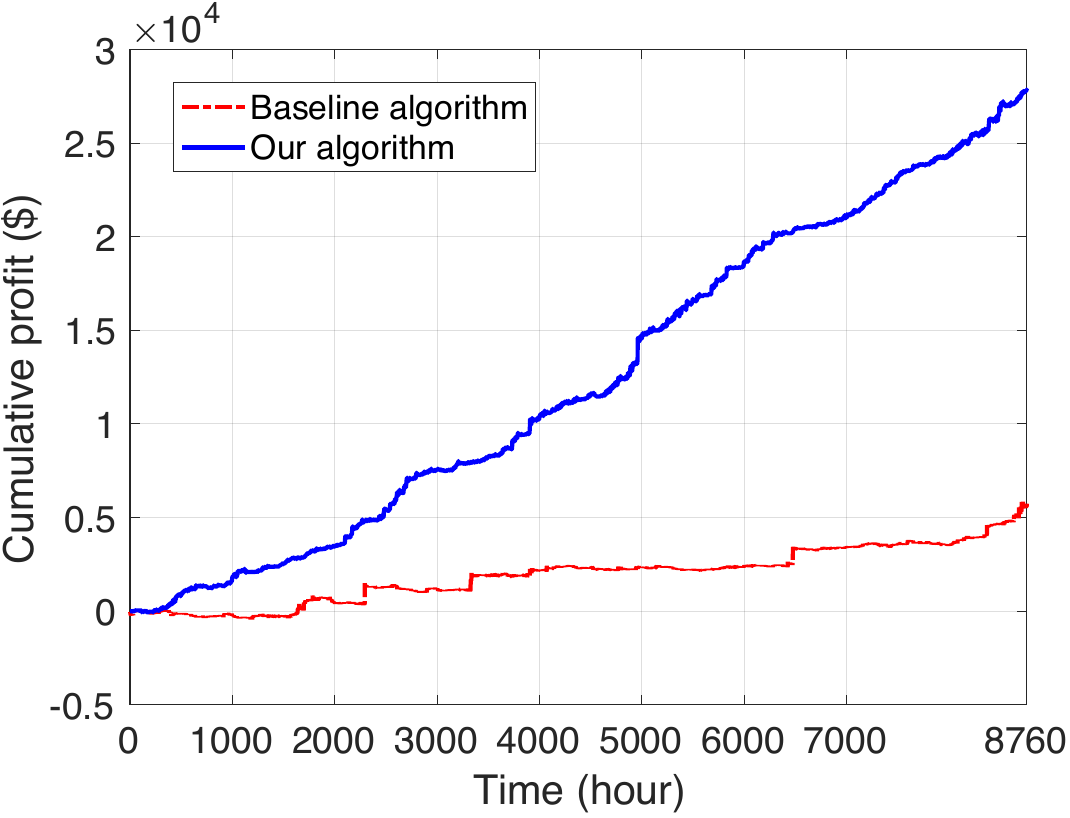}
		\caption*{(a) 8MWh-1MW battery.}
	\end{minipage} 
	\begin{minipage}[!htbp]{0.46\linewidth}
		\centering
		\includegraphics[width=\linewidth]{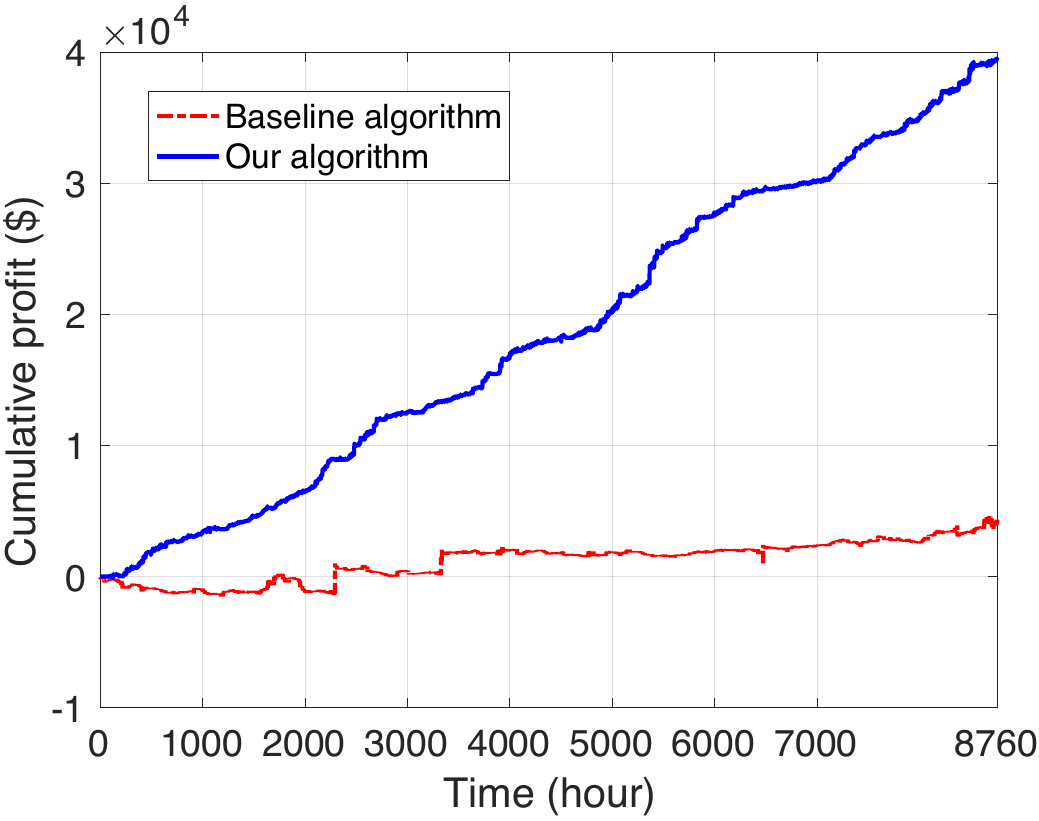}
		\caption*{(b) 8MWh-2MW battery.}
	\end{minipage} 
	\caption{Cumulative profits of the baseline algorithm in~\cite{qin2016online} and our algorithm.}
	\label{fig_profit_baseline}
\end{figure}

Above discussion demonstrates that Reward 2 performs much better than Reward 1, and thus we stick to Reward 2 and compare our algorithm with a baseline algorithm called online modified greedy algorithm in \cite{qin2016online}. This algorithm uses a thresholding strategy to control charge and discharge in an online fashion. We configure the parameters for the baseline according to~\cite{qin2016online}. The arbitrage profits of two algorithms are simulated on an $8-$MWh battery, with a charge/discharge rate of 1MW as depicted in Figure \ref{fig_profit_baseline}. The baseline algorithm can only get $\$5,845$, while our algorithm earns $\$28,027$ that is $4.8$ times of the baseline profit. The profit of the baseline decreases when the charge/discharge rate increases to 2MW. But our algorithm achieves even a higher profit, i.e., $\$39,690$, which is $8.6$ times of the baseline profit $\$4,603$. The reason is that the baseline algorithm relies on the off-line estimate of the price information and lacks adaptability to the real-time prices. Our algorithm updates the average price to adapt to the price changes and thus performs better.

\newpage
\section{Conclusion}\label{conclusion}
In this paper, we derive an arbitrage policy for energy storage operation in real-time markets via reinforcement learning. Specifically, we model the energy storage arbitrage problem as an MDP and derive a Q-learning policy to control the charge/discharge of the energy storage. We design a reward function that does not only reflect the instant profit of charge/discharge decisions but also incorporate the history information. Simulation results demonstrate our designed reward function leads to significant performance improvement and our algorithm achieves much more profit compared with existing baseline method. We will consider self-discharge and degradation of battery in our future work.

\section*{Acknowledgment}
This work was partially supported by the University of Washington Clean Energy Institute.

\bibliographystyle{IEEEtran}	% (uses file "plain.bst")
{\small \bibliography{paperref}}% expects file "myrefs.bib"

\end{document}